\documentclass[12pt]{article}

\usepackage[a4paper,left=2cm,right=2cm,top=2.5cm,bottom=3.5cm]{geometry}

\usepackage{babel}
\usepackage{url} 
\usepackage{graphicx} 
\usepackage{amsmath} 
\usepackage{amssymb}
\usepackage{amsfonts} 
\usepackage{subcaption} 
\usepackage{caption} 
\usepackage{amsthm} 
\usepackage{tocloft} 
\usepackage{xcolor} 
\usepackage{tablefootnote}
\usepackage{soul} 
\usepackage{mathtools} 
\usepackage{cite}

\usepackage[%
pdfsubject={Phase diagram of DNN},
pdfkeywords={Phase diagram},
colorlinks=true,    
citecolor=blue,     
linkcolor=black,    
urlcolor=blue,      
]{hyperref}

\graphicspath{ {./images/} }

\newcommand{\be}{\begin{equation}}
\newcommand{\ee}{\end{equation}}
\newcommand{\bea}{\begin{eqnarray}}
\newcommand{\eea}{\end{eqnarray}}

\newcommand{\qqquad}{\qquad\qquad}
\newcommand{\half}{\frac{1}{2}}

\newcommand{\ket}{\rangle}
\newcommand{\bra}{\langle}

\newcommand{\cD}{{\cal D}}
\newcommand{\cH}{{\cal H}}
\newcommand{\cL}{{\cal L}}
\newcommand{\cN}{{\cal N}}
\newcommand{\cB}{{\cal B}}
\newcommand{\cS}{{\cal S}}
\newcommand{\nn}{\nonumber}
\newcommand{\E}{\mathbb{E}}
\newcommand{\V}{\mathbb{V}}

\newcommand{\Norm}[1]{\left\lVert #1 \right\lVert}

\newcommand{\arctanh}{\mbox{arctanh}}

\newcommand{\svar}[2][]{\mathbb{V}_{#1} \left[ #2 \right]}
\newcommand{\sexpv}[2][]{\mathbb{E}_{#1} \left[ #2 \right]}

\newcommand{\updated}[1]{{\color{black}{#1}}}

\renewcommand{\theequation}{\arabic{section}.\arabic{equation}}
\counterwithin*{equation}{section}

\begin{document}

\title{\bf\large
Phase diagram and eigenvalue dynamics of stochastic gradient descent in multilayer neural networks
}

\author{
Chanju Park$^{a}$,
Biagio Lucini$^{b}$, and Gert Aarts$^{a}$ \footnote{Emails: chanju.b.park@gmail.com, b.lucini@qmul.ac.uk, g.aarts@swansea.ac.uk}
 \\
\mbox{} \\
{\small 
${}^a$Centre for Quantum Fields and Gravity} \\
{\small Department of Physics, Swansea University, Swansea SA2 8PP,  United Kingdom} \\
{\small
${}^b$School of Mathematical Sciences, Queen Mary University of London} \\
{\small
London E1 4NS, United Kingdom} \\
}

\date{November 18, 2025}

\maketitle

\vspace*{-0.5cm}

\begin{abstract}
  Hyperparameter tuning is one of the essential steps to guarantee the convergence of machine learning models. We argue that intuition about the optimal choice of hyper\-parameters for stochastic gradient descent can be obtained by studying a neural network's phase diagram, in which each phase is characterised by distinctive dynamics of the singular values of weight matrices.
  Taking inspiration from disordered systems, we start from the observation that the loss landscape of a multilayer neural network with mean squared error can be interpreted as a disordered system in feature space, where the learnt features are mapped to soft spin degrees of freedom, the initial variance of the weight matrices is interpreted as the strength of the disorder, and temperature is given by the ratio of the learning rate and the batch size. 
  As the model is trained, three phases can be identified, in which the dynamics of weight matrices is qualitatively different. Employing a Langevin equation for stochastic gradient descent, previously derived using Dyson Brownian motion, we demonstrate that the three dynamical regimes can be classified effectively, providing practical guidance for the choice of hyperparameters of the optimiser. 
\end{abstract}


\newpage

\tableofcontents

\hypersetup{linkcolor=blue}   

\section{Introduction}
\label{sec:motivation}

With the discovery of large machine learning (ML) models and their capability of generalisation \cite{zhang2017understandingdeeplearningrequires}, 
significant activity has developed in adopting ML methods in the physical sciences
\cite{Carleo:2019ptp}, including in quantum physics
\cite{Dawid:2022fga} and (lattice) quantum chromodynamics
\cite{Boyda:2022nmh,Cranmer:2023xbe,Aarts:2025gyp}.
From the perspective of a physicist, ML systems are intriguing to study in their own right in the context of statistical physics of complex systems \cite{zdeborova_understanding_2020}.
Indeed, a deeper understanding of ML architectures as systems with many fluctuating degrees of freedom, evolving ``out of equilibrium'' during the training phase, may shed light on how and why certain systems are highly successful. 

A description of learning rooted in statistical physics dates back to (at least) the Hopfield model \cite{1672070, doi:10.1073/pnas.79.8.2554, Hopfield2,PhysRevA.32.1007}, 
which effectively captures the information storage aspect of the neural network \cite{Gardner_1988}, demystifying the scalability and generalisability of large models \cite{K_Nakanishi_1997, PhysRevE.96.042156, Albanese2024HebbianLF}. However, it does not necessarily include practical training settings, which play a crucial role in the performance of a model.
In general, it turns out that deep neural networks during training can be thought of as disordered systems in a non-equilibrium setting with non-Gaussian random couplings, for which analytical solutions are not easily accessible. Some recent work has suggested that these systems can be studied by considering an ensemble of models at a fixed training time \cite{Erbin_2022,Demirtas:2023fir}, where the non-Gaussianity can be mitigated by a perturbative approach, or in the limit of infinite width \cite{lee2018deep, NEURIPS2018_5a4be1fa, JMLR:v22:20-1123,Roberts_2022,10.1214/23-AAP1933}, where the relevant distributions become Gaussian.

In this paper, we suggest that multilayer neural networks,  including the training stage, can be studied by observing the training dynamics of the weight matrices, whose singular values undergo Dyson Brownian motion \cite{Aarts:2024wxi} and follow qualitatively different dynamics depending on the choice of hyperparameters.
We start by recalling that the loss function of a multilayer neural network can be interpreted as a disordered Hamiltonian \cite{PhysRevA.32.1007, Decelle_2021}, where we further argue that the variance of the weight matrices can be interpreted as the strength of the disorder, and the stochasticity of the training induces the notion of effective temperature \cite{Yaida2018FluctuationdissipationRF, JMLR:v18:17-214, Granziol2020LearningRA, Mandt2015ContinuousTimeLO, Aarts:2024wxi}.
Then, depending on the hyperparameters, three different ``phases'' of the training dynamics can be observed, where each phase is characterised by distinctive weight matrix dynamics.
Such phase diagrams of ML architectures are expected to provide theoretical insight into the capacity of the model and the choice of hyperparameters \cite{schoenholz2017deep, bassi2025leftfootleadsright, doi:10.1073/pnas.2316301121, JMLR:v22:20-1123, doi:10.1073/pnas.2311810121, PhysRevE.96.042156, d'amico2025pseudolikelihood, achilli2025capacitymodernhopfieldnetworks}.
Here, we study the trainability of the model depending on the ratio of the learning rate and batch size of stochastic gradient descent, and the initial variance of the weight matrices, which provides guidance to the optimal choice of hyperparameters with regard to the training dynamics.
  
This paper is structured as follows.
In Sec.~\ref{sec:disordered} we define a multilayer neural network and explain how it can be interpreted as a disordered system. We discuss in detail how the degrees of freedom and hyperparameters in the neural network are mapped to ``soft spins'' and physical parameters (temperature, disorder) in the disordered system.
In Sec.~\ref{sec:phase} we present the empirical phase diagram of a neural network with two hidden layers and hyperbolic tangent activation functions. In particular, we analyse a range of observables and identify three phases during and after training. The phases depend on the choice of hyperparameters and are identified with an ordered or ferromagnetic phase, in which the network learns well, a disordered or jamming phase, in which the network does not learn, and a paramagnetic phase, in which the dynamics is dominated by fluctuations preventing learning. 
In Sec.~\ref{sec:dynamics}, we focus on the singular values of the weight matrices and demonstrate that they follow qualitatively different dynamics in each phase, closely related to the existence of stationary distributions at the end of training. To characterise the phase structure further, we derive expressions for the phase boundaries, using a symmetry-breaking argument as well as a stochastic equation for the average level spacing.
In the final section, we show good agreement between these phase boundaries and the regions in the empirical phase diagram obtained by numerical simulations, and discuss the practical implications for successful learning.  
Some more details on the interpretation of the loss function and on the average level spacing can be found in two appendices.

\section{Deep neural networks as disordered systems}
\label{sec:disordered}

\subsection{Feed-forward neural network and loss function}

To make the connection between neural neural networks and disordered systems, we start by defining the former, following closely the notation of Ref.~\cite{Roberts_2022}. We consider a neural network with $L+1$ layers, with each layer consisting of $n_l$ nodes ($l=0,\ldots, L$). The first layer is the input layer, with the input data given by $n_0$-dimensional vectors, and the final layer is the output layer, with $n_L$ components.  
The layers are connected by $L$ weight matrices $W^{(l)}$ ($l=1, \ldots, L$) of size $n_l\times n_{l-1}$. We do not include a bias (but this can easily be done).
The input data set is indicated as
\be
\cD = \{ x_{i\alpha}\} \qqquad i=1, \ldots, n_0, \qquad \alpha=1, \ldots, |\cD|,
\ee
where the first index ($i$) is the component index and the second index ($\alpha$) labels each sample in the data set. Activation functions, denoted as $\phi(z)$, are applied at the hidden nodes and act component-wise. Here, the $z$'s are the pre-activations, i.e., linear combinations of the components of the previous layer. Explicitly, the first and subsequent pre-activations are then given by
\be
z_i^{(l+1)}(x_\alpha) = \sum_{j=1}^{n_{l}} W_{ij}^{(l+1)} \phi \left(z_{j}^{(l)} (x_\alpha)\right),
  \qqquad 
  z_i^{(1)}(x_\alpha) = \sum_{j=1}^{n_{0}} W_{ij}^{(1)} x_{j\alpha},
\ee
and the result on the final layer defines the neural network function
\be
\label{eq:nn}
    \hat y_i(x_\alpha; \theta) \equiv z_i^{(L)}(x_\alpha) = \sum_{j=1}^{n_{L-1}} W_{ij}^{(L)}  \phi \left(z_{j}^{(L-1)} (x_\alpha)\right).
\ee 
This function depends on the $N_\theta$ learnable parameters, collectively denoted as  
\be
\theta = \{W^{(1)}, \ldots, W^{(L)}\}, \qqquad 
N_\theta = \sum_{l=1}^L n_{l-1}n_l.
\ee
The choice of activation function is discussed below. 
The outputs of the final hidden layer are often called features or representations learnt by the neural network \cite{6472238}, such that the neural network function is a linear combination of these features. Below we indicate the features as
\be
\phi_{j\alpha} \equiv \phi \left(z_{j}^{(L-1)} (x_\alpha)\right).
\ee
In this paper, we consider the mean squared error (MSE) as the loss function, i.e.,
\be
\label{eq:loss1}
    \cL \left( \theta \right) \equiv \frac{1}{|\cD|}
    \sum_{\alpha=1}^{|\cD|} \ell \left( y(x_\alpha),
    \hat y(x_\alpha; \theta) \right),
    \qqquad
    \ell \left( y, \hat y \right) \equiv \frac{1}{2} \sum_{i=1}^{n_L} \left(
    y_i - \hat y_i \right)^2,
\ee
where $\ell$ is the per-sample loss function for input data sample $x_\alpha \in\cD$, with target value $y_i(x_\alpha)\equiv y_{i\alpha}$. 

To make the connection with disordered systems, we expand the loss function and write
 \begin{align}
    \cL \left( \theta \right) &= \frac{1}{2|\cD|} \sum_{\alpha=1}^\cD \sum_{i=1}^{n_L} \left( y_{i\alpha}  - \sum_{j=1}^{n_{L-1}} W_{ij}^{(L)}\phi_{j\alpha} \right)^2 \nn \\
    &=
    \frac{1}{2|\cD|}\sum_{\alpha=1}^{|\cD|} \sum_{i=1}^{n_L}
    \left( 
    \sum_{j,k=1}^{n_{L-1}} W^{(L)}_{ij} W^{(L)}_{ik} \phi_{j\alpha} \phi_{k\alpha} 
    - 2 \sum_{j=1}^{n_{L-1}} y_{i\alpha}    W_{ij}^{(L)} \phi_{j\alpha} 
    +  y_{i\alpha} y_{i\alpha}  
    \right) \nn \\
    &= 
    \frac{1}{2|\cD|} \sum_{\alpha=1}^{|\cD|} \sum_{i,j=1}^{n_{L-1}} J_{ij} \phi_{i\alpha} \phi_{j\alpha} 
    - \frac{1}{|\cD|} \sum_{\alpha=1}^{|\cD|} \sum_{j=1}^{n_{L-1}} h_{j\alpha} \phi_{j\alpha} 
    + C.
  \end{align}  
where to reach the final line, we renamed the indices and introduced
  \begin{align}
 \label{eq:parameters}
    J_{ij} \equiv \sum_{k=1}^{n_L} W^{(L)}_{ki} W^{(L)}_{kj},
    \qquad
    h_{j\alpha} \equiv \sum_{i=1}^{n_L} y_{i\alpha} W_{ij}^{(L)},
    \qquad
    C \equiv \frac{1}{2|\cD|} \sum_{\alpha=1}^{|\cD|} \sum_{i=1}^{n_L} y_{i\alpha}^2.
  \end{align}
Here we intentionally have chosen a notation that resembles the one familiar from disordered systems, with the features $\phi_i$ playing the role of {\em spin} degrees of freedom, interacting via a {\em spin-spin coupling} $J_{ij}$ and with an {\em external magnetic field} $h_j$, see also Appendix \ref{sec:duality}. The final term is independent of the neural network and will be dropped; hence, from now on, we consider the loss function
 \begin{align}
    \cL \left( \theta \right) =  \frac{1}{2|\cD|} \sum_{\alpha=1}^{|\cD|} \sum_{i,j=1}^{n_{L-1}} J_{ij} \phi_{i\alpha} \phi_{j\alpha} 
    - \frac{1}{|\cD|} \sum_{\alpha=1}^{|\cD|} \sum_{j=1}^{n_{L-1}} h_{j\alpha} \phi_{j\alpha}. 
      \label{eq:loss}
  \end{align}  
\updated{Note that this form of mean squared error is generic for any neural network model whose output is defined as a linear combination of the nodes on the last hidden layer, regardless of the structure of the preceding layers.}
In the remainder of this section, we further explore the relation to disordered systems. To do so, we remind the reader that the usual spin-glass Hamiltonian is written for binary spins, $s_i=\pm 1$, as \cite{PhysRevLett.35.1792}
\be
\label{eq:disorder}
\cH = -\half \sum_{i,j} J_{ij}s_is_j + \sum_j h_js_j,
\ee
where $J_{ij}$ is the random coupling and $h_j$ is a random external field, both drawn from a Gaussian distribution. Phase diagrams of spin-glass systems have been studied since the 1970s, and may exhibit ferromagnetic, paramagnetic and spin-glass phases, in the plane spanned by temperature and the strength of the disorder. 

To analyse our loss function (\ref{eq:loss}) in the language of disordered systems and arrive at a phase diagram, we now address the various ingredients, namely the initalisation of the network, the feature degrees of freedom $\phi_{i\alpha}$, disorder and the coupling $J_{ij}$, the external field $h_{j\alpha}$, and the emergence of an effective temperature when using stochastic gradient descent.

\subsection{Initialisation}

The network is initialised by sampling the weight matrix elements from normal distributions, according to 
\be
\label{eq:Winit}
W^{(l)}_{ij} \sim \cN\left(0, \sigma_W^2/n_{l-1}\right),
\ee
where we will refer to $\sigma_W^2$ as the weight matrix variance (the factor $n_{l-1}$ is discussed below). In principle, weight matrices connecting different layers can have their own variance, $\sigma_W^2\to \sigma_W^{(l)2}$, but for notational simplicity we take them identical.
Denoting the pre-activations as
\be
z_{i\alpha}^{(l)} \equiv z_{i}^{(l)}(x_\alpha),
\ee
moments of pre-activations at initialisation over the weight matrix distributions can be com\-put\-ed recursively \cite{NIPS2016_14851003,Roberts_2022}. The first moments vanish for symmetry reasons. The second moments read, for $l>1$, 
\begin{align}
\E_{p(W)}\left[ z_{i\alpha}^{(l)} z_{j\beta}^{(l)}\right] 
& = \sum_{k,k'=1}^{n_{l-1}} \E_{p(W)}\left[ W_{ik}^{(l)} W_{jk'}^{(l)} \right] 
\phi\left(z_{k\alpha}^{(l-1)}\right) \phi\left( z_{k'\beta}^{(l-1)}\right)
\nn \\
& =
\delta_{ij} \frac{\sigma_W^2}{n_{l-1}}  \sum_{k=1}^{n_{l-1}} 
\phi\left(z_{k\alpha}^{(l-1)}\right) \phi\left( z_{k\beta}^{(l-1)}\right),
\end{align}
while at the first layer, one finds
\be
\E_{p(W)}\left[ z_{i\alpha}^{(1)} z_{j\beta}^{(1)}\right] 
= \sum_{k,k'=1}^{n_0} \E_{p(W)}\left[ W_{ik}^{(1)} W_{jk'}^{(1)} \right] 
x_{k\alpha} x_{k'\beta}
=
\delta_{ij} \frac{\sigma_W^2}{n_{0}}  \sum_{k=1}^{n_0}  x_{k\alpha} x_{k\beta}.
\ee
Averaging also over the data set, assuming it is standardised, with 
\be
\E_\cD\left[ x_{i\alpha}x_{j\beta}\right] = \delta_{ij} \delta_{\alpha\beta},
\ee
then yields
\be
\E_{p(W), \cD}\left[ z_{i\alpha}^{(1)} z_{j\beta}^{(1)}\right] = \delta_{ij} \delta_{\alpha\beta}\sigma_W^2.
\ee
The important observation for us is that all second moments scale with $\sigma_W^2$. 

As activation function we use the hyperbolic tangent, $\phi(z)=\tanh(z)$, which is bounded between $\pm 1$. We then trivially have 
\be
\left| \phi\left(z_{k\alpha}^{(l)}\right) \phi\left( z_{k'\beta}^{(l)}\right) \right| \leq 1,
\ee
and hence find
\be
\left|\E_{p(W)}\left[ z_{i\alpha}^{(l)} z_{j\beta}^{(l)}\right] \right| \leq  \frac{\sigma_W^2}{n_{l-1}}  \sum_{k=1}^{n_{l-1}} 
\left| \phi\left(z_{k\alpha}^{(l-1)}\right) \phi\left( z_{k\beta}^{(l-1)}\right) \right| \leq \sigma_W^2.
\ee
At initialisation, the variances of the pre-activations therefore scale with and are bounded by $\sigma_W^2$; this is one reason to normalise the weight matrix variance with $n_{l-1}$. Note that the scaling with $\sigma_W^2$ always holds, while the boundedness is due to the boundedness of the activation function.

\subsection{Features as ``soft spins''}
\label{sec:softspins}

\begin{figure}[tp!]
 \centering
 \includegraphics[width=0.9\linewidth]{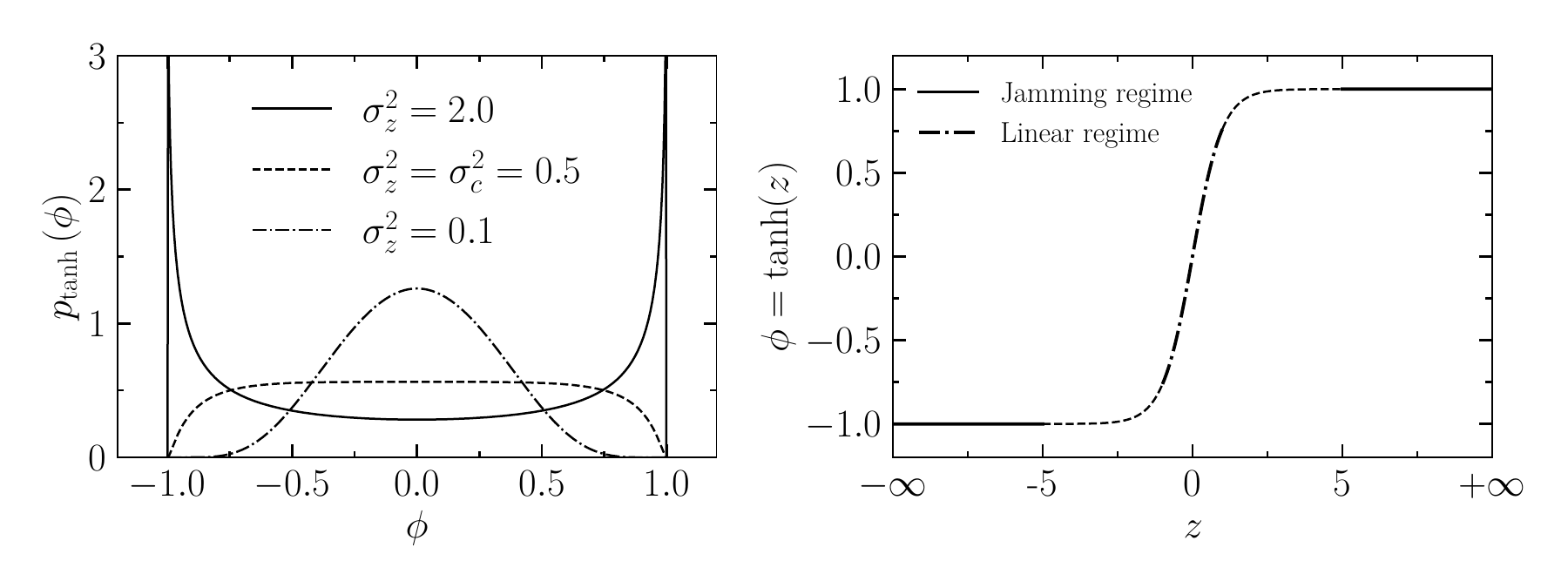}
 \caption{Left: Distribution of post-activation $\phi(z)$, with $z\sim \cN(0, \sigma_z^2)$. For small $\sigma_z^2<1/2$, the distribution is peaked around zero, while for large $\sigma_z^2\gg 1/2$, the distribution is sharply peaked towards $\pm 1$.
 Right: Hyperbolic tangent function $\phi(z)=\tanh(z)$ as activation function, with a linear regime ($|z|\lesssim 1$) and a jamming or vanishing gradient regime  ($|z| \gg 1$).
 }
 \label{fig:tanh}
\end{figure}

Disordered systems are usually formulated in terms of binary spins, $s_i=\pm 1$, see Eq.~(\ref{eq:disorder}). Here we argue that the features, $\phi_{i\alpha}$, play the role of ``soft spins'', which take continuous values, but are still bounded between $\pm 1$, due to the choice of the hyperbolic tangent activation function.
The distribution of features at initialisation depends on pre-activations, $z_{j\alpha}^{(L-1)}$. 
As shown above, the variance of the pre-activation scales with (and is bounded by) $\sigma_W^2$. 
Motivated by results in the large width limit of neural networks \cite{Roberts_2022} and the central limit theorem, we assume that 
\be
z^{(L-1)}_{i\alpha} \sim \cN(0,\sigma^2_z), \qqquad\sigma_z^2 \lesssim \sigma_W^2.
\ee
Dropping the indices for simplicity, we write
\be
\phi(z) = \tanh(z), \qqquad -1\leq \phi(z) \leq 1,
\ee
with the distribution
\be
p(\phi) \sim p(z(\phi)) \left|\frac{dz}{d\phi} \right| \sim \exp\left[ - \frac{1}{2\sigma_z^2}\left(\arctanh\phi\right)^2 - \ln(1-\phi^2) \right] \sim \exp[-V(\phi)].
\ee
Expanding the potential around $\phi=0$, we find
\be
V(\phi) =  \left( \frac{1}{2\sigma_z^2} -1\right) \phi^2 + \left( \frac{1}{3\sigma_z^2} -\half \right) \phi^4 +\ldots,
\ee
which exhibits a transition from a single well to a double well at $\sigma_z^2=\sigma_c^2=1/2$. This observation also holds without the expansion, as the extrema of the potential are determined by $\phi=\tanh(2\sigma_z^2\phi)$. 
We conclude that the distribution of features has a single peak centred around $0$ when $\sigma_z^2\leq 1/2$ and a double peak when $\sigma_z^2>1/2$, as demonstrated in Fig.~\ref{fig:tanh} (left). 

The activation function is shown in Fig.~\ref{fig:tanh} (right). We can now draw
a first conclusion on how the initial variance of the weight matrices, a hyperparameter, affects the learning efficiency for bounded activation functions.
For large $\sigma_z^2\sim \sigma_W^2$, features $\phi$ are close to $\pm 1$. The activation function is then mostly in the vanishing gradient regime, and poor learning is expected. On the other hand, for small $\sigma_z^2\sim \sigma_W^2$, the features lie around 0, which corresponds to the linear regime in the activation function, where good learning is expected. This vanishing gradient transition can be regarded as a jamming transition in disordered systems, where a distribution of pre-activations $z$, with variance $\sigma_z^2$, is placed in a finite box (the activation function) with a vanishing gradient on the boundary, and the dynamics of the distribution starts to get jammed once the size of the distribution exceeds a critical value $\sigma_c^2$. Below, we will numerically confirm that indeed a large initial variance of the weight matrices results in poor learning, corresponding to a jammed phase.

\subsection{Disorder}

Next, we turn to the coupling between the features, $J_{ij}$, the product of the weight matrices at the final layer, 
\be
J_{ij} = \sum_{k=1}^{n_L} W^{(L)}_{ki} W^{(L)}_{kj},
\ee
see Eq.\ (\ref{eq:parameters}). 
In a standard disordered system with $N$ spins \cite{PhysRevLett.35.1792}, the spin couplings are drawn from a normal distribution, $J_{ij}\sim \cN(J_0, J^2)$, and a transition from a ferromagnetic phase to a spin-glass phase is observed as $\sqrt{N}J_0/J$ is reduced, where the scaling with $N$ ensures a proper thermodynamic limit. 

In the case considered here, at initialisation the matrix elements of $W^{(L)}$ are drawn from a normal distribution with variance $\sigma_W^2/n_{L-1}$. The $n_{L-1}\times n_{L-1}$ coupling matrices $J_{ij}$ are hence symmetric and positive semi-definite, and belong to a Wishart-Laguerre ensemble with Dyson index $\beta = 1$ \cite{livan_introduction_2018}. In the limit of large matrix size, its spectral density tends to the Marchenko-Pastur distribution defined on a finite support.

At initialisation, the mean and covariance of the $J$'s are given by
\begin{align}
\E_{p(W)}\left[ J_{ij}\right] = \frac{n_L}{n_{L-1}} \sigma_W^2 \delta_{ij}, 
\qquad
\mbox{Cov}\left[ J_{ij} J_{kl}\right]  
= \frac{n_L}{n_{L-1}^2} \sigma_W^4 \left(
\delta_{ik} \delta_{jl} + \delta_{il} \delta_{jk}
\right).
\end{align}
To determine what variable to vary to observe a possible transition between an ordered and a disordered phase, we may follow the choice in the disordered spin system and use mean divided by $\sqrt{\mbox{variance}}$, $J_0/J$. However, for the Wishart ensemble, this scales as $\sigma_W^0$ and hence it is not a suitable combination. Instead, we will use the scaling with $1/\sqrt{\mbox{variance}}$ of the initial weight matrices, i.e., $1/\sigma_W$. As we will demonstrate below, this choice indeed provides a useful control parameter.

\subsection{External field alignment}
\label{sec:ext}

The second term in the loss function (\ref{eq:loss}) is the coupling between the features $\phi_{j\alpha}$ and the {\em external magnetic field} $h_{j\alpha}$,
\be
 \cL_h(\theta) = - \frac{1}{|\cD|} \sum_{\alpha=1}^{|\cD|} \sum_{j=1}^{n_{L-1}} h_{j\alpha} \phi_{j\alpha}, 
\qqquad  
 h_{j\alpha} = \sum_{i=1}^{n_L} y_{i\alpha} W_{ij}^{(L)},
\ee
where $y_{i\alpha}$ is the target value and both $h_{j\alpha}$ and $\phi_{j\alpha}$ depend on the input data.

Given that the prediction of the network is
\be
  \hat y_{i\alpha} = \sum_{j=1}^{n_{L-1}} W_{ij}^{(L)}  \phi_{j\alpha}, 
\ee 
this term can also be written as
\be
 \cL_h(\theta) = - \frac{1}{|\cD|} \sum_{\alpha=1}^{|\cD|} \sum_{i=1}^{n_L} y_{i\alpha} \hat y_{i\alpha}
 \ee
which is minimised when the output vectors are aligned, as expected.
This explains the relation between the alignment of the output data on one hand and the features and external field on the other hand.

\subsection{Temperature}
  
An important difference between disordered systems and neural networks is that the former are usually considered at a temperature $T$, whereas the latter evolve during training, from initialisation to potentially being well-trained, and should therefore be considered as non-equilibrium systems. 
Indeed, weight matrix elements are typically drawn from normal dis\-tri\-bu\-tions at initialisation and training will induce correlations, leading to characteristic heavy-tailed distributions of weight matrix spectra \cite{DBLP:journals/corr/abs-1810-01075,pmlr-v97-mahoney19a}, known from random matrix theory \cite{PhysRevB.110.L180102, akemann_power-law_2008}.

Nevertheless, it is possible to identify a notion of temperature, characterising fluctuations observed during training. This can be made precise in the case of stochastic gradient descent (SGD), one of the most commonly used training algorithms. Each weight matrix element is updated according to 
\begin{align}
\label{eq:W}
 W_{ij}^{(l)\prime} = W_{ij}^{(l)} - \epsilon \Delta_{ij, \cB}^{(l)},
 \qquad
 \Delta_{ij, \cB}^{(l)} = \frac{1}{|\cB|}\sum_{\alpha\in\cB}  \Delta_{ij, \alpha}^{(l)}
 \qquad
    \Delta_{ij, \alpha}^{(l)} = \frac{\partial \ell(x_\alpha)}{\partial W_{ij}^{(l)}}.
\end{align}
Here $\epsilon > 0$ is the learning rate (or step size), which we take fixed in this analysis, $ \Delta_{ij, \alpha}^{(l)}$ is the element-wise gradient of the loss function for data point $x_\alpha$, and updates are performed using batches of size $|\cB|$. 

Stochasticity is induced because of the finite batch size. The mean and fluctuations of the gradient can be separated using the central limit theorem, assuming the input data is i.i.d. The evolution of the model parameters can then be described by a discrete Langevin equation,
\begin{align} 
\label{eq:langevin}
 W_{ij}^{(l)\prime} = W_{ij}^{(l)} - \epsilon \sexpv{\Delta_{ij}^{(l)}}
 + \sqrt{\frac{\epsilon^2}{|\cB|} \svar{\Delta_{ij}^{(l)}}} \, \eta_{ij},
 \qquad
 \eta_{ij} \sim \mathcal{N} \left(0, 1\right).
\end{align}
Here $\sexpv{\Delta}$ and $\svar{\Delta}$ are the mean and variance of the gradient, respectively, and the scaling with learning rate and batch size has been derived in Ref.~\cite{Aarts:2024wxi} in the context of Dyson Brownian motion.
The scaling with the step size does not follow the standard (It\^{o}) scaling in (discretised) stochastic differential equations (SDEs). In Ref.~\cite{Aarts:2024wxi} it was demonstrated that the particular scaling in Eq.~(\ref{eq:langevin}) leads to the linear scaling rule, i.e., a dependence on the ratio $\epsilon/|\cB|$ only.\footnote{The dependence on the ratio $\epsilon/|\cB|$, and not on the hyperparameters $\epsilon$ and $|\cB|$ separately, has been observed empirically in SGD optimisation and is known as the linear scaling rule \cite{l.2018dont,goyal2018accuratelargeminibatchsgd}.
It implies that decreasing the learning rate by a factor $k$ has the same effect as increasing the batch size by the same factor. This is explained by noting that the effective temperature $T=\epsilon/|\cB|$ is invariant under a simultaneous scaling of both.} 
Moreover, this ratio combines with the Coulomb potential in the random matrix theory description of Dyson Brownian motion in the same way as a temperature would \cite{Aarts:2024wxi}. 

An alternative way to arrive at $\epsilon/|\cB|$ as an effective temperature is by taking the limit of zero step size in a weak sense. A naive $\epsilon\to 0$ limit would lead to deterministic gradient flow without stochasticity
  \cite{Yaida2018FluctuationdissipationRF, JMLR:v18:17-214, ziyin_parameter_2024}.
Instead, the correct SDE limit is obtained if the ratio $\epsilon/|\cB|$ is kept fixed when the learning rate $\epsilon$ and the batch size $|\cB|$ are taken to zero simultaneously. This yields a Langevin equation in continuous time, 
\begin{align}
\label{eq:langevint}
 \dot{W}^{(l)}_{ij} = - \sexpv{\Delta_{ij}^{(l)}}
    + \sqrt{T \svar{\Delta_{ij}^{(l)}}} \, \eta_{ij},
    \qqquad
 T\equiv\frac{\epsilon}{|\cB|},
      \end{align}
where the dot indicates the time derivative and the effective temperature $T$ is a measure of the stochasticity, appearing as a temperature in the stationary solution of the corresponding Fokker-Planck equation (assuming that it exists) \cite{Aarts:2024wxi}.  
Below, we identify $\epsilon/|\cB|$ with the temperature axis in the phase diagram, to which we turn now.

\section{Empirical phase diagram}
\label{sec:phase}

In the preceding section, we have argued that a neural network with hyperbolic tangent activation functions and subject to the MSE loss function should be considered as a disordered system, with a temperature set by the ratio of learning rate and batch size, $T=\epsilon/|\cB|$, disorder by the variance $\sigma_W^2$ of the weight matrices upon initialisation, and features $\phi_i$ as soft spins. In contrast to a disordered spin system at a given temperature, the learnable parameters in a neural network, i.e.\ the weight matrix elements, evolve during training, and our aim is to determine how the efficiency and accuracy of training dynamics change depending on the temperature and initial variance. A practical application of this study is that it provides guidance on how to choose the hyperparameters. 

As argued above, we consider the phase diagram in the plane spanned by the temperature and the inverse square root of the initial variance, the $T-1/\sigma_W$ plane.
Intuitively, three characteristic phases resembling a disordered system \cite{PhysRevLett.35.1792,doi:10.1142/0271} can be expected: 
\begin{itemize}
    \item a high-temperature or paramagnetic phase, in which the model is subject to large fluctu\-ations and the features of the model are randomly distributed;
    \item a disordered spin-glass or jamming phase at large $\sigma_W$, in which the model parameters accumulate near gradient vanishing points and the training dynamics is jammed from initialisation;
    \item a low-temperature or ferromagnetic phase with little disorder, in which the model is capable to learn the target features and reach an equilibrium state at the end of training.
\end{itemize}
Traditionally, in statistical mechanics, a phase diagram is studied by identifying symmetries and order parameters, and subsequently calculating the system's free energy. Here, the neural network is evolving during training, which calls for a non-equilibrium treatment, and the identification of order parameters is not obvious. Hence, we probe the phase diagram empirically, using numerical experiments, 
and consider observables which are both inspired by disordered systems but also relevant in the context of learning, such as the loss, the gradient of the loss, and alignment at the end of training.

\subsection{Numerical setup}

 We consider a neural network with two hidden layers ($L=3$): the input layer has $n_0 = 3$ input neurons, the hidden layers have $n_1 = 32$ and $n_2 = 16$ neurons respectively, and the output is a scalar, $n_{L=3} = 1$. The task is regression and, as stated, the activation functions are hyperbolic tangents.
Specifically, we use a teacher-student setting, where the training data for the network are generated from a teacher neural network with weight matrices $\overline{W}^{(l)}$ ($l=1,2,3$). The matrix elements in the teacher network, as well as the input data, are sampled from a normal distribution,
  \begin{align}
    \overline{W}_{ij}^{(l)} \sim \cN(0, 1/n_{l-1}),
    \qqquad
    x_{i\alpha} \sim \cN(0, 1).
  \end{align}
  Note that the (scaled) variance in the teacher weight matrices is equal to 1. 
 This setup determines the output of the teacher network. 
  
The student network is defined with the same architecture as the teacher network, but with the intermediate weight matrix $W^{(2)}$ set to be trainable and initialised according to Eq.~(\ref{eq:Winit}). The first and final student weight matrices are identical to the corresponding teacher ones, $W^{(1)}= \overline{W}^{(1)}$, $W^{(3)}= \overline{W}^{(3)}$. 
This setting can be regarded as a variation of a Random Feature Neural Network \cite{rahimi_random_2007}, where instead of given random features, the linear combinations are given and the features representing the correct target feature space are learnt.
From a theoretical perspective, this setting enables us to isolate the dynamics of a single layer and carry out a precise numerical study of the phase diagram and weight matrix dynamics.
\updated{Moreover, using the teacher-student setting is useful as one has direct control over the target distribution for the student network, allowing one to check the theoretical predictions with experimental outcomes.}

In the numerical experiments discussed below, for each combination of hyperparameters, an ensemble $\cS$ with a total of $|\cS|=100$ teacher networks is generated, and each student network is trained on a different teacher network.
The hyperparameters are selected as follows: the learning rate is varied from $\epsilon = 2^{-3}, \ldots, 2^8$ in powers of 2, with a batch size of $|\cB|=4$. Hence the temperature range is $T=\epsilon/|\cB| = 2^{-5}, \ldots, 2^6$. The inverse square root of the variance of the student weight matrix $W^{(2)}$ is varied as $1/\sigma_W=2^{-8}, \ldots, 2^3$. Each model is trained with the same number of iterations $t=10^5$.
 A typical training set consists of $|\cD|=4000$ data points.

\subsection{Observables}

We now come to the results of the numerical experiments and introduce four observables that capture different aspects of the dynamics of learning across the phase diagram. The training time is denoted with $t$ and the end of the training with $t=t_f$. All observables are evaluated on a test set with $|\cD_{\rm test}|=100$ data points after training. 

First, we measure the mean test loss at the end of the training, defined as an average of the loss (\ref{eq:loss1}) over 
the ensemble $\cS$ of trained networks at a fixed choice of hyperparameters,  
\begin{align}
 \E_{\cS}[\cL] = \frac{1}{|\cS|} \sum_{s=1}^{|\cS|} \cL^s,
\end{align}
where $\mathcal{L}^s$ is the final test loss of the $s$-th model in the ensemble $\cS$.
A high mean loss indicates that the model has not converged to the correct solution and the training is unsuccessful, while a low mean loss indicates that the model is well trained.

\begin{figure}[t]
    \centering
    \includegraphics[width=0.5\linewidth]{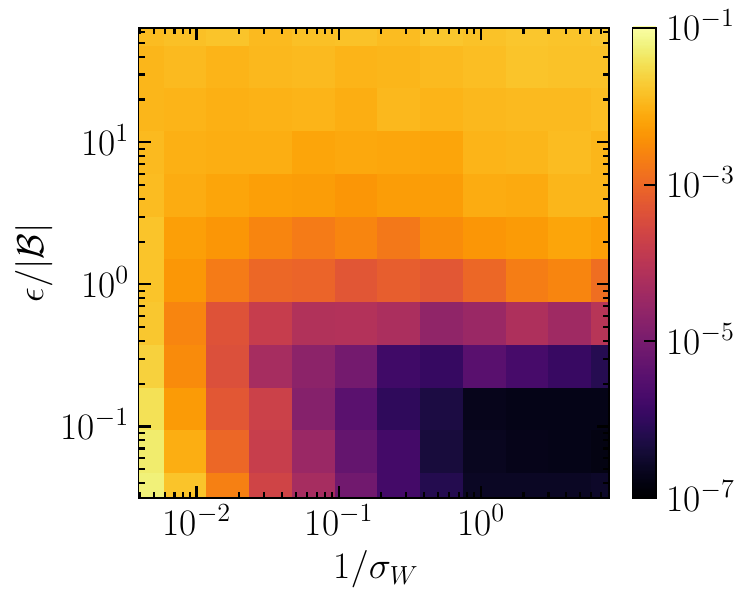}
    \caption{The mean test loss of the trained models in the $T-1/\sigma_W$ plane for the hyperbolic tangent teacher-student network. A darker colour indicates a lower value. 
    }
    \label{fig:obs1}
\end{figure}
  
The results for the final mean loss are shown in Fig.~\ref{fig:obs1}, where the mean loss is seen to vary between approximately $10^{-1}$ and $10^{-7}$ in the plane spanned by $T=\epsilon/|\cB|$ and $1/\sigma_W$. The darker region corresponds to a smaller final loss.
At higher temperatures and at larger initial variance, the loss remains large, indicating that the ensembles of models are not close to the target ones. On the other hand, at low temperature and small initial variance, i.e., in the bottom-right corner of this phase diagram, the average loss is substantially smaller, and we may deduce that the ensembles of models have converged with high probability. 
Below, we identify the bottom-right corner in Fig.~\ref{fig:obs1} with a ferromagnetic phase.

As a second observable, we consider the mean gradient to quantify whether the training is active or dormant. We define the mean gradient as
\begin{align}
  \E_\cS[ \Norm{\nabla \cL}]
    = \frac{1}{|\cS|} \sum_{s=1}^{|\cS|}
    \sqrt{ \sum_{i=1}^{n_2} \sum_{j=1}^{n_1} 
    \left(\Delta^{(2),s}_{ij}\right)^2 },
  \end{align}
where $\Delta^{(2),s}_{ij}$ is the gradient of the loss function for the $s$-th model with respect to $W_{ij}^{(2)}$, averaged over the data set.

 \begin{figure}[t]
    \center
    \includegraphics[width=\linewidth]{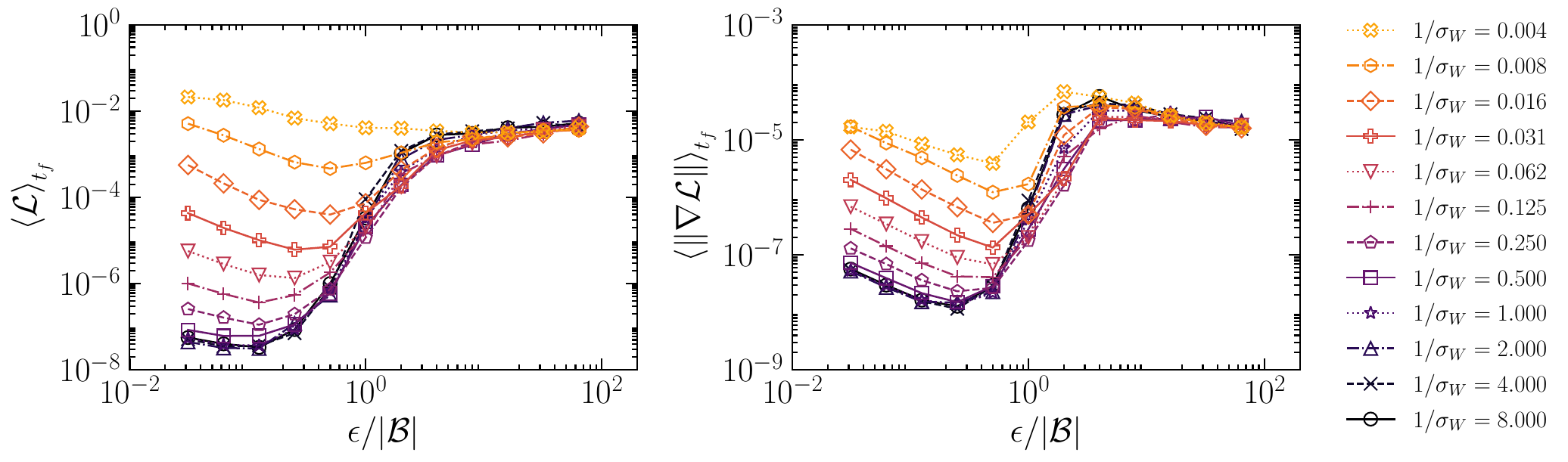}
    \caption{Mean loss (left) and gradient (right) at the end of training as a function of $\epsilon/|\cB|$ for various values of $1/\sigma_W$. Error bars are omitted for visibility.}
    \label{fig:slow}

      \centering
      \includegraphics[width=\linewidth]{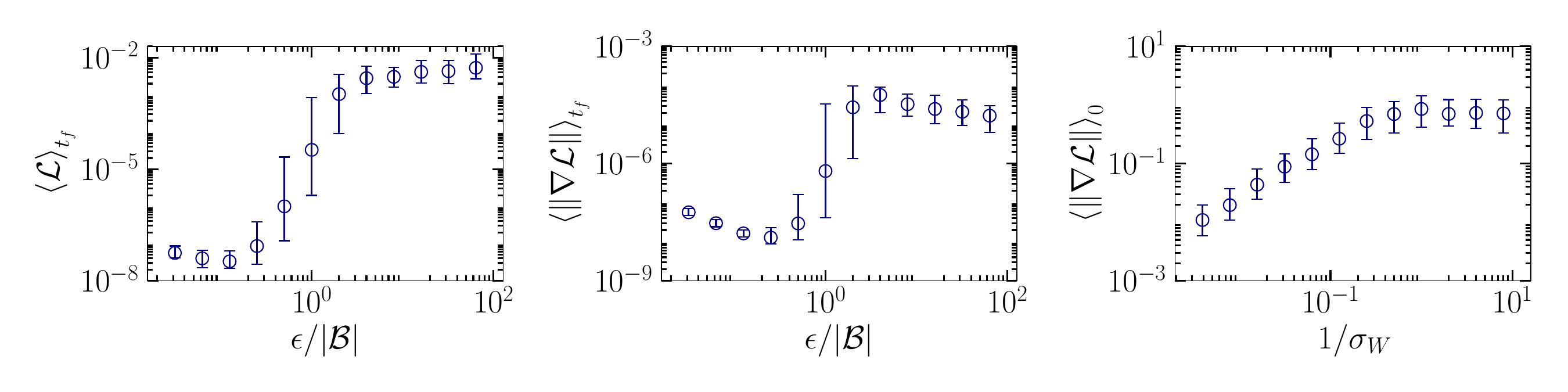}
      \caption{The mean loss (left) and mean gradient (centre) at the end of training along the temperature axis at the smallest value of $\sigma_W$ considered, $1/\sigma_W = 8$. A transition is visible at $\epsilon/|\cB|\sim 1$.
      Mean gradient at initialisation (right) along the $1/\sigma_W$ axis at the lowest temperature, $\epsilon/|\cB| = 2^{-5}$. For small $1/\sigma_W$ (large initial disorder), learning is already inefficient at initialisation.
      }
      \label{fig:TS_loss_grad}
  \end{figure}
  
In Fig.~\ref{fig:slow}, we show the mean loss and gradient at the end of training as a function of $\epsilon/|\cB|$ for various values of $1/\sigma_W$, i.e., each collection of data points connected by a line corresponds to a vertical slice of the preceding phase diagram. Error bars are omitted for visibility, but are shown in Fig.~\ref{fig:TS_loss_grad} for $1/\sigma_W=8$. We observe different behaviour, which we identify with the phases discussed above:

\begin{itemize}

\item If the temperature and initial variance are small, $\epsilon/|\cB| \sim 10^{-1}$, $1/\sigma_W\gtrsim 1$, both the final loss and gradient are small, indicating that the models have learnt successfully and that learning is completed. This corresponds to the ordered or ferromagnetic phase.

\item Increasing the initial variance at small temperature results in a larger loss, while the gradient remains large as well.  This indicates that learning remains active but without making progress.  This is identified with a jamming phase with slow dynamics. %

\item At higher temperature, $\epsilon/|\cB| \sim 10^{0}$, a transition is observed to a regime where
both the final loss and final gradient are large, indicating a lack of convergence. Since this dynamics is independent of the matrix initialisation, it signifies a transition to a paramagnetic phase.\footnote{
\updated{This corresponds to the dotted regions in the phase diagrams analysed in Ref.~\cite{doi:10.1073/pnas.2316301121}, which are considered in the plane spanned by the learning rate and the batch size.} }

\item At even higher temperature, $\epsilon/|\cB| \sim 10^{2}$, and for all values of $1/\sigma_W$, the dependence on the initialisation has vanished and the models are clearly in the paramagnetic phase, with the dynamics dominated by thermal fluctuations.

\end{itemize}
In Fig.~\ref{fig:TS_loss_grad}, we add error bars for the loss (left) and the gradient (middle) as a function of $\epsilon/|\cB|$ for the smallest value of $\sigma_W$ considered, $1/\sigma_W = 8$. We note a rapid increase at $\epsilon/|\cB|\sim 1$, in a manner that is not dissimilar to a phase transition, albeit in a finite volume or with a finite number of degrees of freedom. 
In  Fig.~\ref{fig:TS_loss_grad} (right), we show the gradient of the loss at initialisation, along the $1/\sigma_W$ axis at the lowest temperature considered.  
We observe that for large initial variance, $1/\sigma_W  \sim 10^{-2}$, the mean gradient is already small at initialisation, resulting in slow spin-glass-like dynamics, while with small initial variance, $1/\sigma_W \gtrsim 10^0$, the dynamics is much more efficient.
We argue, therefore, that for practical implementations, the optimal values of the learning rate over batch size are located right before the transition from the low-temperature to the high-temperature phase, with small initial variance for the weight matrix elements, allowing for fast training while yielding good convergence.

\begin{figure}[t]
    \centering
    \includegraphics[width=0.96\linewidth]{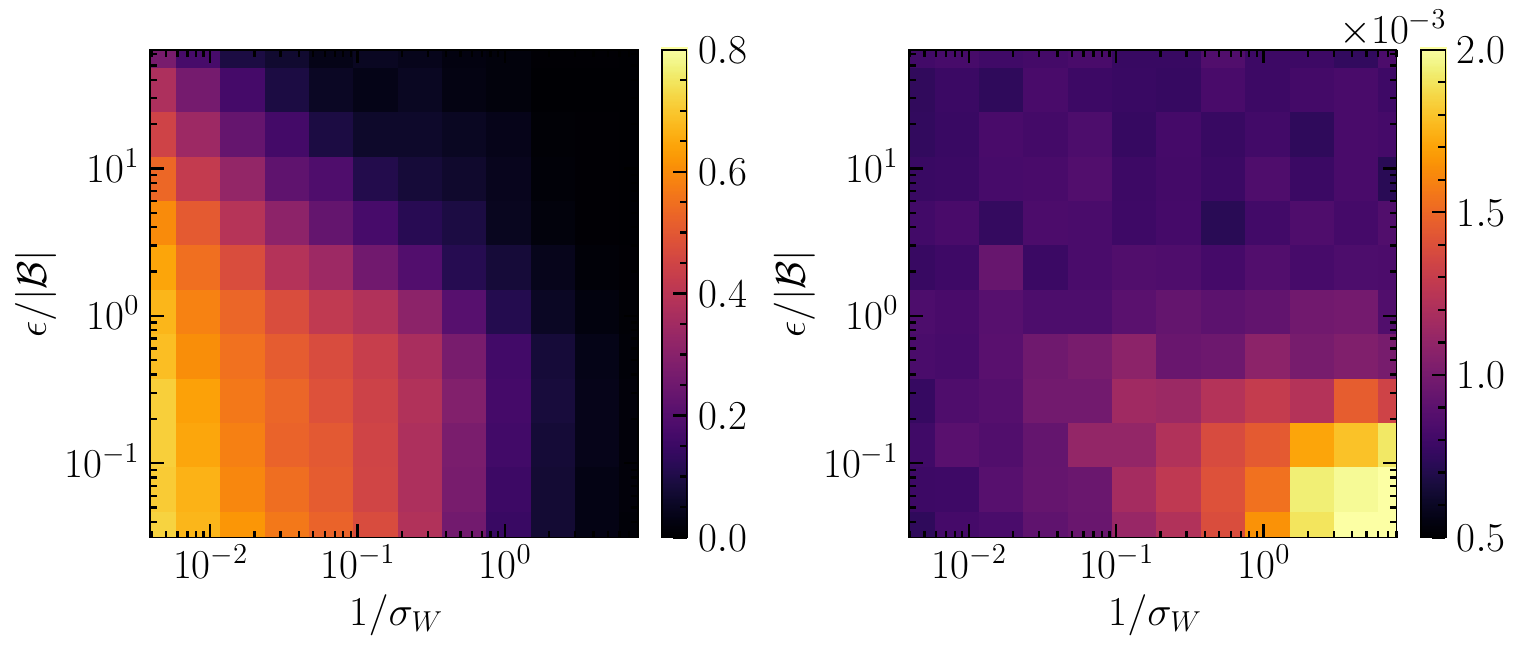}
    \caption{The correlation between features at initial and final time, $G(t_f,0)=\E[\phi(t_f) \phi(0)]$, (left) and 
    between the features $\phi$ and the external magnetic field $h$, $\E[h \phi]$, or alignment between model and target features, (right), in the $T-1/\sigma_W$ plane for the hyperbolic tangent teacher-student network. A darker colour indicates less correlations.}
    \label{fig:obs2}
  \end{figure}

The next observable we consider is the time correlation function of features, defined by 
  \begin{align}
    G(t,t') \equiv \E_\cS[\phi(t) \phi(t')] = \frac{1}{|\cS|} \sum_{s \in \mathcal{S}}  \frac{1}{|\cD|} \sum_{\alpha \in \mathcal{D}}
    \frac{1}{n_{L-1} } \sum_{j=1}^{n_{L-1}}  \phi_{j\alpha}^{s} (t) \phi_{j\alpha}^{s} (t'),
  \end{align}
  where $\phi^s_{j\alpha}(t)$ is a component of a feature at time $t$ in the $s$-th model. 
We consider in particular the correlation between the initial features, at $t=0$, and the features at the end of training, at $t=t_f$, i.e., $G(t_f,0)$.   
The results for $G(t_f,0)$ are shown in Fig.~\ref{fig:obs2} (left). A darker colour corresponds to a lower value and hence less correlations, whereas a lighter colour implies that correlations are preserved. The observed correlations in Fig.~\ref{fig:obs2} are in agreement with the discussion presented so far. In the high-temperature (paramagnetic) phase and in the well-trained (ferromagnetic) phase, the correlations between features at initialisation and after training are lost, as expected. With a large initial variance, on the other hand, correlations are preserved, which is interpreted as poor learning in the disordered spin-glass phase. Increasing the temperature reduces this correlation, but leads to poor training due to the transition to the paramagnetic phase.

The final observable we consider is the alignment between target and model features, via the correlation between the features $\phi_{j\alpha}$  and the external magnetic field $h_{j\alpha}$, see Sec.\ \ref{sec:ext}. We define
   \begin{align}
 \E_\cS[ h \phi] = \frac{1}{|\cS|} \sum_{s=1}^{|\cS|} \frac{h^s \cdot \phi^s}{\Norm{h^s} \Norm{\phi^s} }
  \end{align}
  where 
   \be
    h^s\cdot\phi^s = \sum_{\alpha=1}^{|\cD|} \sum_{j=1}^{n_{L-1}} h_{j\alpha}^{s} \phi_{j\alpha}^{s},
\qqquad
  \Norm{h^s}^2 = \sum_{\alpha=1}^{|\cD|} \sum_{j=1}^{n_{L-1}} h_{j\alpha}^{s} h_{j\alpha}^{s},
 \ee
 and similar for $\Norm{\phi^s}^2$.
 The results are shown in Fig.~\ref{fig:obs2} (right). We observe the strongest alignment in the well-trained region of the phase diagram, giving a clear justification to associate the region with the ferromagnetic phase.

In summary, we have considered several observables in the plane spanned by the hyper\-pa\-ram\-e\-ters. Each observable probes different characteristics of learning, allowing us to identify the various phases. The behaviour of the mean loss gives the first indication of a well-trained phase. The alignment with the external magnetic field emphasises the similarity with a ferromagnetic phase. The difference between the paramagnetic phase and the disordered spin-glass phase is captured by the correlations of features in time, which are preserved in the latter. These findings are consistent with the understanding of different phases in disordered systems, using the mapping between the hyperparameters and the parameters in generic spin-glass models.

\section{Dynamics of training}
\label{sec:dynamics}

In this section, we take a complementary approach, i.e., rather than focussing on the features, we study the dynamics of the weight matrices and demonstrate that the various phases can also be observed in the evolution of these. We consider the $n_2\times n_1$ weight matrix $W^{(2)}$, where in our case $n_2=16<n_1=32$. Following Ref.~\cite{Aarts:2024wxi}, we consider the symmetric  combination 
\be
X = W^{(2)} W^{(2)\,T},
\ee
and denote its eigenvalues with $x_i$ ($i=1,\ldots, n_2$). These eigenvalues are semipositive and are the squares of the singular values of $W^{(2)}$.

\begin{figure}[t]
    \centering
    \includegraphics[width=\linewidth]{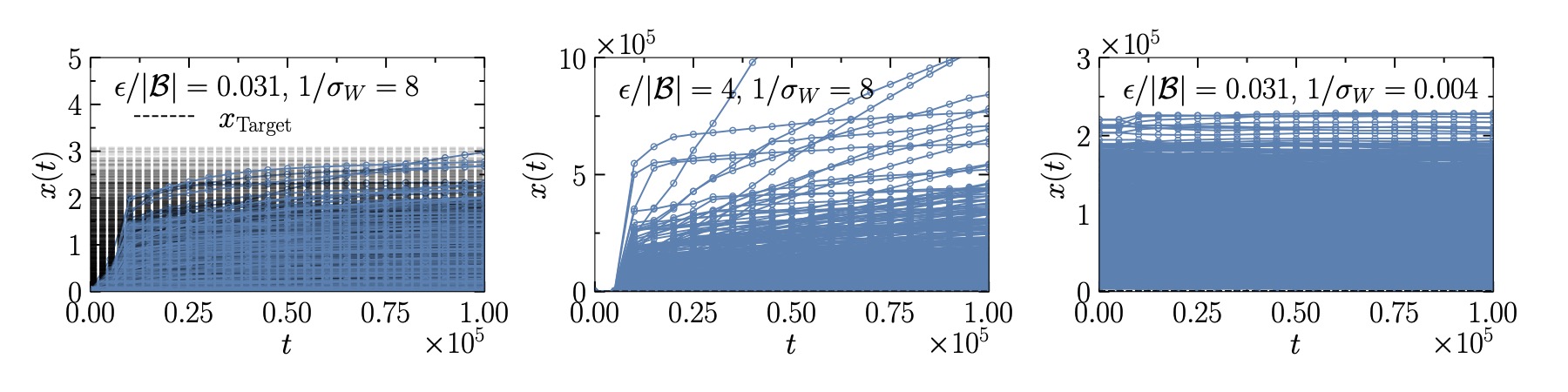}
    \caption{Evolution of the eigenvalues of $X = W^{(2)}W^{(2)\,T}$ during training for ensembles of teacher-student models at three choices of hyperparameters, 
    in the ordered phase (left) with $\epsilon/|\cB|=2^{-5}$, $1/\sigma_W=8$, 
    in the high-temperature phase (centre) with $\epsilon/|\cB|=4$, $1/\sigma_W=8$, 
    and in the disordered  phase (right) with $\epsilon/|\cB|=2^{-5}$, $1/\sigma_W=2^{-8}$. 
    Dashed horizontal lines denote the target eigenvalues and are only visible on the left. Note the difference in scale on the vertical axis.
    }
    \label{fig:eig_flow}
  \end{figure}

The evolution of the eigenvalues $x_i(t)$ during training is presented in Fig.~\ref{fig:eig_flow} in the ordered phase (left), the high-temperature phase (centre), and the disordered phase (right). Dashed horizontal lines denote the target eigenvalues and are only visible on the left. Note the difference in scale on the vertical axis. What is shown in each figure is the evolution for an ensemble of 100 teacher-student models, hence a total of $1600$ eigenvalues. 

At initialisation, with $W^{(2)}_{ij}\sim \cN(0, \sigma_W^2/n_1)$, the distribution of eigenvalues of $X$ is given by the Marchenko-Pastur distribution,
\be
\label{eq:MP}
P_{\rm MP}(x) = \frac{1}{2\pi\sigma_W^2 rx}\sqrt{(x_+-x)(x-x_-)}, \qqquad x_-<x<x_+,
\ee
where $r=n_2/n_1=1/2$, $x_\pm=\sigma_W^2(1\pm \sqrt{r})^2$. The upper limit is hence given by $x_+=0.046$ for $1/\sigma_W=8$ and $x_+=1.9\times 10^5$ for $1/\sigma_W=2^{-8}$. This explains the difference in the initial range between the figure on the right and the two other examples. 

We observe dynamics characteristic of all three phases. In the ordered phase (left), the distribution of eigenvalues flows towards the target distribution. In the high-temperature phase (centre), stochastic fluctuations are so strong that the distribution quickly widens and the eigenvalues grow until they potentially reach a gradient vanishing point. In the jamming phase (right), the initial distribution is already very broad, and eigenvalues evolve slowly without ever converging.

\subsection{Stochastic dynamics of the average level spacing}

The dynamics of the eigenvalues can in principle be modelled by generalised Dyson Brownian motion \cite{Aarts:2024wxi}, derived from the stochastic equation for the weight matrix, with discrete updates (\ref{eq:langevin}) or in continuous time (\ref{eq:langevint}).
Dyson Brownian motion for the eigenvalues of $X$ in continuous time, with $T=\epsilon/|\cB|$, is given by \cite{Aarts:2024wxi, dyson_brownian-motion_1962, mehta_random_1967},
\begin{align} 
  \label{eq:dyson}
    \dot{x}_i
    = K_i + T \sum_{j \neq i}   \frac{V_{ij}}{x_i - x_j}  + \sqrt{T V_{ii}} \, \eta_i,
    \qqquad
    \eta_i \sim \mathcal{N}(0, 1).
\end{align}
Here $K_i=K_{ii}$ and $V_{ii}$ are the diagonal elements of the mean and variance of the gradient after diagonalisation of the former, and $V_{i\neq j}$ are the off-diagonal components, with
  \begin{align}
    K_{ij} = -\E[\Delta_{ij}^{X}],
    \qquad
    V_{ij} = \V[\Delta_{ij}^{X}],
    \qquad
    \Delta_{ij}^{X} = \sum_{k=1}^{n_1} \left( W_{ik}^{(2)} \Delta_{kj}^{(2)} + \Delta_{ik}^{(2)} W_{kj}^{(2)} \right),
  \end{align}
 where the final equation follows from the update for $W^{(2)}$, see Eq.~(\ref{eq:W}). The second term in Eq.~(\ref{eq:dyson}) is the Coulomb term, leading to eigenvalue repulsion.

\begin{figure}[t]
    \centering
    \includegraphics[width=0.5\textwidth]{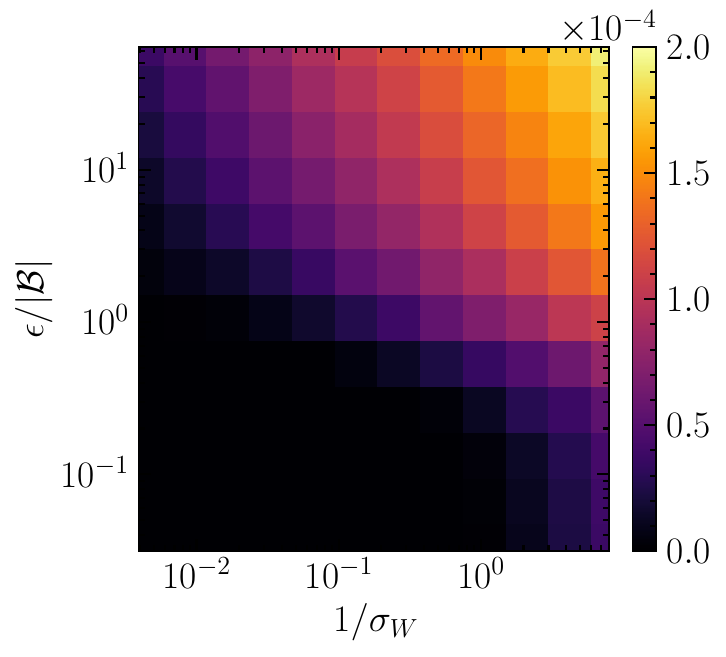}
    \caption{Time-averaged rate at the end of learning, $\overline\lambda(t_f)$, in the $T-1/\sigma_W$ plane. A darker colour means a smaller value. 
    }
    \label{fig:St}
  \end{figure}

Solving the coupled set (\ref{eq:dyson}) of SDEs would give the time evolution of the eigenvalues, but it is not straightforward to do so, as the drift, Coulomb term, and diffusion coefficients are complicated non-linear and time-dependent functions.
To simplify the problem, we introduce a ``one-particle theory", see also Ref.~\cite{PhysRevLett.95.246101}, by focussing on the average level spacing.
Let $x_i$ be the $i$-th eigenvalue of $X$ and assume that the eigenvalues are ordered, $0\leq x_1 < x_2 < \cdots < x_{n_2}$. We consider the level spacing $S_i = x_{i+1}  - x_i$, and its average value,
\begin{align}
\label{eq:ls}
 S = \frac{1}{n_2-1} \sum_{i=1}^{n_2-1} S_i = \frac{1}{n_2-1} \left( x_{n_2} - x_1 \right).
\end{align}
As a first Ansatz, we assume that the average level spacing effectively evolves with a local rate $\lambda(t)$, i.e.,
\be
\dot S(t) = \lambda(t) S(t)
\qquad
\Rightarrow
\qquad 
S(t) = S(0) e^{\int_0^t ds\, \lambda(s)}.
\ee
We can then define a time-averaged rate, 
\be
\overline\lambda(t) = \frac{1}{t} \int_0^t ds\, \lambda(s)
=\frac{1}{t} \log\frac{S(t)}{S(0)}.
\ee
The computed values at the end of learning, $\overline\lambda(t_f)$, are shown in Fig.~\ref{fig:St}.
We observe that the averaged rate is smallest in the jammed phase and largest in the paramagnetic phase, consistent with the time correlation of the features shown in Fig.~\ref{fig:obs2} (left).\footnote{In the paramagnetic phase, at the highest temperature and smallest variance and at $t_f\sim 10^5$, see Fig.~\ref{fig:eig_flow} (centre), we find $(n_2-1)S(t_f)\sim 10^6$, $(n_2-1)S(0) = x_+-x_-\sim 0.04$, and hence $\overline\lambda(t_f)\sim 10^{-5}\log(10^6/0.04)\sim 2 \times 10^{-4}$, in agreement with Fig.~\ref{fig:St}.
}

To improve on this, we construct a stochastic equation for $S$, by subtracting the equation for the smallest eigenvalue $x_1$ from the one for $x_{n_2}$, see Eq.~(\ref{eq:ls}). A detailed derivation is given in Appendix~\ref{sec:derivation}. This results in
\begin{align} 
 \label{eq:s_langevin}
 \dot{S} = K_S + V_S \frac{T}{S} + \sqrt{2TD_S}\, \eta, \qqquad \eta\sim \cN(0,1).
\end{align}
Expressions for $K_S, V_S$, and $D_S$ in terms of $K_{ij}$ and $V_{ij}$ are given in Eqs.~(\ref{eq:VS}, \ref{eq:KS}).
In the following, we use Eq.~(\ref{eq:s_langevin}) to derive a relation for the phase boundary between the high-temperature phase and the other phases.

\subsection{Phase boundary from a stability analysis}

In the high-temperature phase, the level spacing keeps increasing, see Fig.~\ref{fig:eig_flow} (centre), unlike in the two other phases in which it approximately stabilises after the initial stage, albeit for different reasons. 
We now derive a condition of convergence using a stability analysis of the combined force term in Eq.~(\ref{eq:s_langevin}). A restoring force is required to balance the effects of fluctuations due to the noise, c.f.~the fluctuation-dissipation theorem.

We consider the deterministic part of Eq.~(\ref{eq:s_langevin}), 
\begin{align}
 \dot{S} = K_S + V_S \frac{T}{S},
\end{align}
where in general, $S$ will converge if the RHS is negative and diverge if it is positive. Note that $K_S<0$ close to a (local) minimum, when the loss function is minimised. We assume $V_S>0$.
Hence, the convergence condition is given by the inequality,
\begin{align} 
 \label{eq:inequality}
 \frac{T}{S} \le -\frac{K_S}{V_S} \quad \Leftrightarrow \quad {\rm converge},
 \qqquad
 \frac{T}{S} > -\frac{K_S}{V_S}  \quad \Leftrightarrow \quad {\rm diverge}.
  \end{align}
The quantity on the RHS represents the signal-to-noise ratio of the gradient of the loss function. 
The LHS can be related to the initial hyperparameters, as follows.
If the average level spacing keeps increasing, we have 
\be
\label{eq:inequality_s}
S(t_f)>S(t)>S(0) \qqquad \mbox{or} \qqquad  \frac{1}{S(0)} > \frac{1}{S(t)} > \frac{1}{S(t_f)}.
\ee
Combining this with Eq.~(\ref{eq:inequality}) in the case of divergent dynamics then yields the following inequality during training, 
\begin{align} 
 \frac{T}{S(0)} > \frac{T}{S(t)} > -\frac{K_S(t)}{V_S(t)}.
\end{align}
On the other hand, a conservative estimate for converging dynamics is given by
\begin{align}
  \frac{T}{S(0)} < -\frac{K_S(t_f)}{V_S(t_f)}.
\end{align}
The initial eigenvalue distribution of $X$ is given by the Marchenko-Pastur distribution (\ref{eq:MP}). Hence, the average level spacing at initialisation is bounded by
\begin{align}
 S(0) \leq \frac{1}{n_2-1} \left( x_+ - x_- \right)
    = \frac{1}{n_2-1} 4 \sigma_W^2 \sqrt{r},
    \qqquad
    r = \frac{n_2}{n_1}.
  \end{align}
The boundary for divergent dynamics is then approximately given by
\begin{align} 
 \label{eq:convergence_boundary}
    \frac{n_2-1}{4 \sqrt{r}} \frac{T}{\sigma_W^2} >
     -\frac{K_S(t_f)}{V_S(t_f)}.
  \end{align}
The quantities on the LHS are variables determining the network setup, initialisation, and details of the SGD ($n_1, n_2, \sigma_W, T=\epsilon/|\cB|$), whereas the quantity on the RHS indicates the signal-to-noise ratio of the gradient at the end of training. 
Eq.~(\ref{eq:convergence_boundary}) therefore establishes 
that the average level spacing converges to a stationary value only when the signal-to-noise ratio of the gradient on the RHS is larger than the stochasticity on the LHS.

  \begin{figure}[t]
    \centering
    \includegraphics[width=0.96\linewidth]{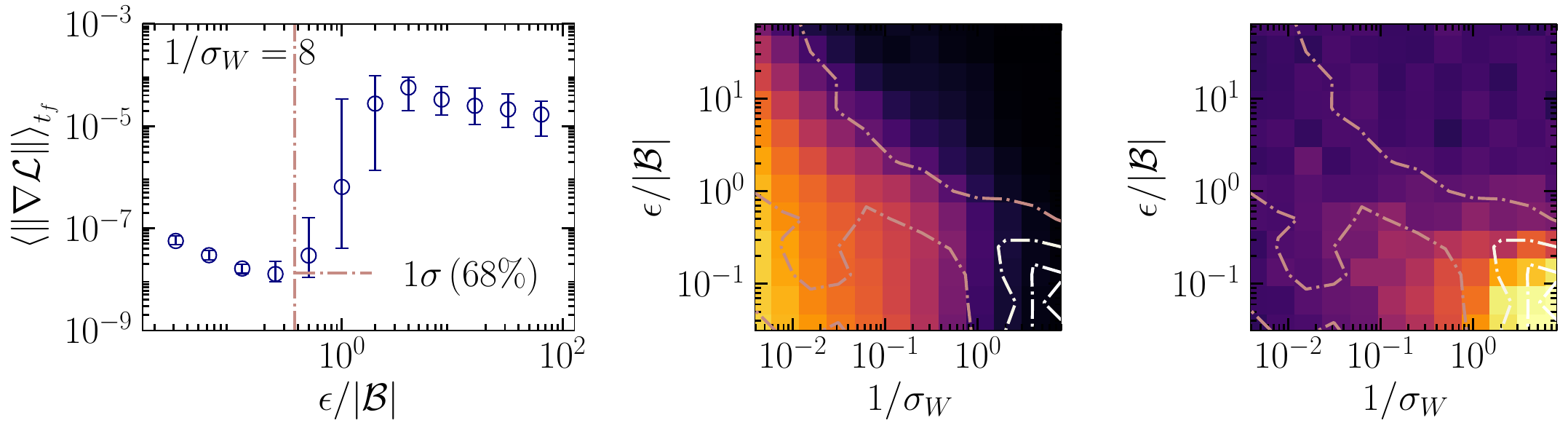}
    \caption{Convergence boundaries obtained from Eq.~(\ref{eq:convergence_boundary}). 
    Left: the vertical dashed-dotted line is the $1\sigma$ boundary at fixed $1/\sigma_W=8$, separating the ordered and the high-temperature phase. 
    Three convergence boundaries on top of the time correlation (centre) and feature alignment (right).  
    The dimmest line indicates that $1\sigma$ $(68\%)$ of the models have converged, and
    the brighter two lines indicate that $2\sigma$ $(95\%)$ and $3\sigma$
    $(99.7\%)$ of the models have converged.
    }
    \label{fig:pred}
  \end{figure}

To apply Eq.~(\ref{eq:convergence_boundary}) in practice, we compute the gradient and variance of the loss function at the end of training and estimate the RHS for each run, see Appendix \ref{sec:derivation} for details. We then compare both sides of Eq.~(\ref{eq:convergence_boundary}) to determine whether, according to this criterion, the dynamics has converged or not. Each point in the phase diagram is assessed using an ensemble of $\cS=100$ models, and we determine whether $1\sigma$ $(68\%)$, $2\sigma$ $(95\%)$, and $3\sigma$ $(99.7\%)$ of the models satisfy the inequality (\ref{eq:convergence_boundary}). 
These boundaries are shown in Fig.~\ref{fig:pred}.
Fig.~\ref{fig:pred} (left) shows the $1\sigma$ boundary for the smallest $\sigma_W$: the vertical line separates the ordered phase and the high-temperature phase, indicating the temperature of the transition. 
In Fig.~\ref{fig:pred} (centre and right) three lines are shown, indicating that 68\%, 95\%, and 99.7\% of the models satisfy inequality (\ref{eq:convergence_boundary}) at the end of the training. In the upper region, above the $1\sigma$ line, the dynamics has not converged due to thermal fluctuations; this is the high-temperature phase. 
This boundary tracks the transition region especially well for the time correlation function, shown in the centre. 
In the lower region towards the left, below the $1\sigma$ line, the dynamics has not converged due to disorder; this is the jamming phase in which the dynamics is slow but not completely static.
Increasing the requirement for convergence, we observe that the $2\sigma$ and $3\sigma$ boundaries capture the ordered phase in the lower-right region very well: here essentially all models converge due to the strong alignment of features.

\section{Discussion and outlook}

\label{sec:conclusion}

Starting from the observation that the loss function of a multilayer neural network can be interpreted as a Hamiltonian of a disordered system, we analysed in some detail its training dynamics under stochastic gradient descent in the case of a teacher-student model. We have shown that three phases can be identified, depending on the choice of hyperparameters. These findings can be conveniently summarised in a phase diagram, in which the ratio of the learning rate over batch size, $T=\epsilon/|\cB|$, determines the  ``temperature axis'', while the initial variance $\sigma_W^2$ of the weight matrix elements sets the ``disorder axis''.

\begin{figure}[t]
    \centering
    \includegraphics[width=0.9\linewidth]{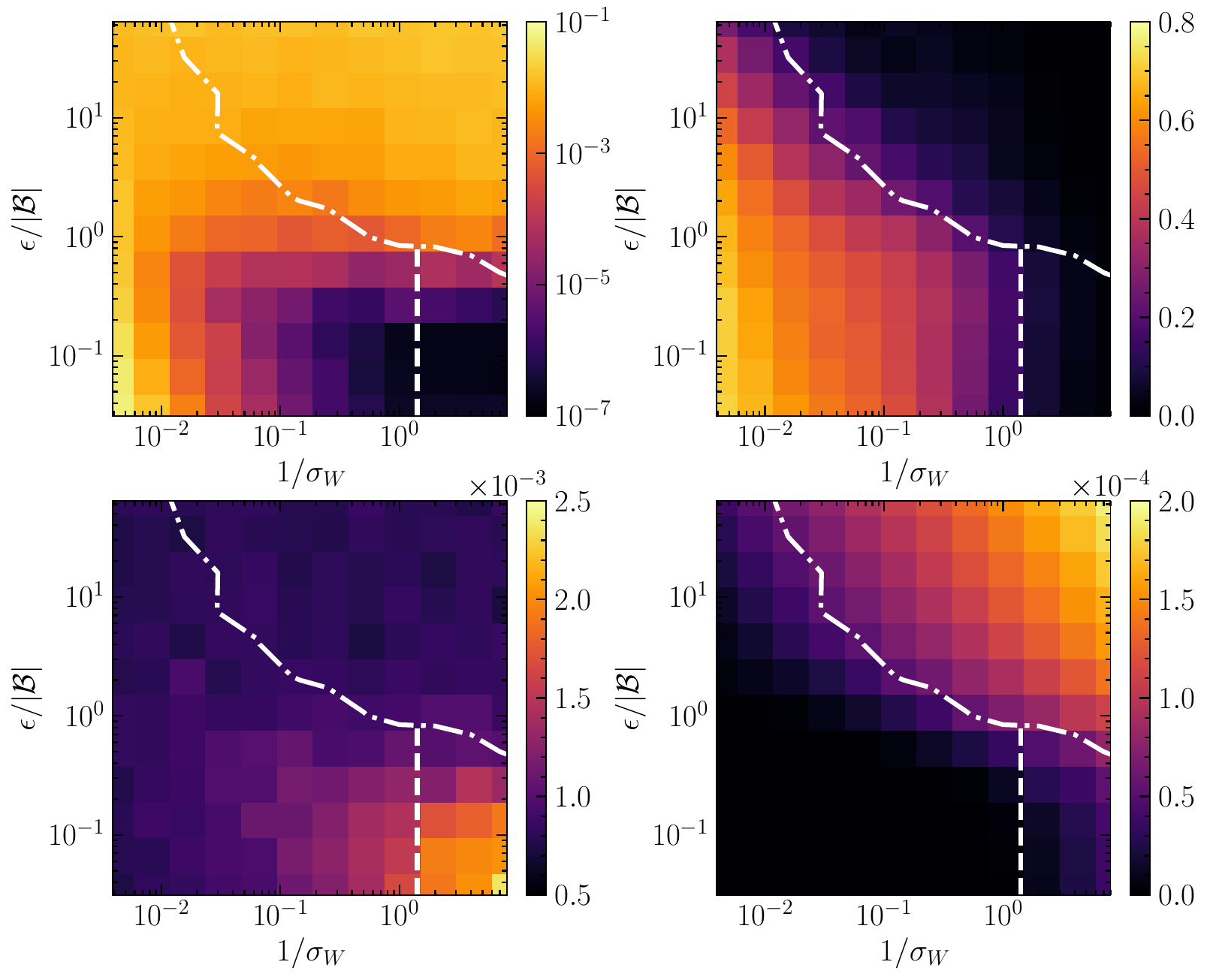}
    \caption{Four observables probing different aspects of the phase structure:
    final test loss (top left), time correlation of features (top right), feature alignment (lower left), and time-averaged rate of the average level spacing  (lower right) in the $T-1/\sigma_W$ plane, with superimposed the estimates of the 
     phase boundary between the jamming and ferromagnetic phase (dashed vertical line) and between the paramagnetic and the other phases (dashed dotted line).
     }
    \label{fig:TS_ran_obs_all}
  \end{figure}

Fig.~\ref{fig:TS_ran_obs_all} contains a summary of our findings, with four observables probing different aspects of the phase structure. Superimposed are the estimates of the phase boundaries between the three phases. Starting in the lower-left corner, the dynamics is dominated by the large initial variance, leading to a jamming phase familiar from spin-glass like systems, in which training is extremely slow. This results in a large loss, poor feature alignment, and persistent memory of the initial state of the network. Reducing the initial variance while keeping the temperature low leads to a transition to a well-trained -- or ferromagnetic -- phase, in which the loss is small, the features are aligned with the target data, and memory of the initial state is successfully erased. The phase boundary is given by the vertical dashed line, which is derived from a symmetry-breaking argument, directly linked to our choice of hyperbolic tangents as activation functions, see Sec.~\ref{sec:softspins}. Increasing the level of stochasticity brings us to the high-temperature -- or paramagnetic -- phase, in which the loss is again large and feature alignment is absent, but due to the fluctuations, the memory of the initial state is erased rapidly. The phase boundary, indicated by the dashed-dotted line, is derived using the framework of Dyson Brownian motion for stochastic weight matrix dynamics and follows essentially a signal-to-noise argument, with the high-temperature phase dominated by noise.

A practical implication of this study is that the phase boundaries can guide the choice of optimal hyperparameters: selecting the largest temperature in the ordered phase leads to fast training. It is noted that to determine the phase boundary to the high-temperature phase, one has to compute the signal-to-noise ratio of the gradient at the end of the training, but having a grasp of what the phase diagram looks like will help in finding optimal hyperparameters effectively, even without explicit knowledge of this boundary.
More generally, the significance of this work lies in the fact that it incorporates stochastic training dynamics in the analysis, proposes a direct interpretation of hyperparameters as physical quantities, and provides intuition from theory on how hyperparameters affect the training dynamics.

There are several directions to explore in the future. The phase boundary between the ordered and disordered phases at low temperature is closely linked to hyperbolic tangent activation functions being bounded. In contrast, activation functions of the ReLU type are only bounded on one side, which will affect the phase structure. This is worth exploring further. In this study, we used a teacher-student model and focused on one of the weight matrices in the neural network. It is of interest to extend this to include realistic data sets \updated{such as MNIST and analyse the dynamics of all weight matrices.} 
Our neural network is relatively small; considering deeper and especially wider architectures may lead to more pronounced phase boundaries and potentially the use of finite-size scaling techniques to analyse the transitions at a more quantitative level. 
\updated{To achieve this, one first needs to define the correct thermodynamic limit, which preserves the dynamical properties of the optimisation, as naive large width or depth limits lead to lazy training, where the norm of the gradient vanishes \cite{NEURIPS2019_ae614c55}.}

While we followed the evolution of the singular values of the weight matrices during training, we did not study the evolution of their spectral density \cite{Aarts:2024wxi,Aarts:2024qey}. Empirical evidence shows that correlations between the weight matrix elements are induced during training, such that the ensemble of weight matrices at the end of training deviates from the Wishart-Laguerre distribution at initialisation, leading, e.g., to heavy tails 
\cite{DBLP:journals/corr/abs-1810-01075,pmlr-v97-mahoney19a}. It would be interesting to analyse these distributions using techniques developed in the random matrix community \cite{akemann_power-law_2008,PhysRevB.110.L180102}.

Once the spectral properties of the trained network and a consistent thermodynamic limit are obtained, a theoretical calculation of the phase boundaries might be feasible using methods developed for disordered systems \cite{guerra_broken_2003, agliari2018novelderivationmarchenkopasturlaw}.
An ambitious long-term objective would then be to study universality properties of different neural network architectures,  potentially leading to insights on designing efficient architectures and predicting scaling behaviour from first principles.

\vspace*{0.2cm} 

\noindent
{\bf Acknowledgements} --  
We thank Ouraman Hajizadeh and Matteo Favoni for discussion.
The main part of this work was carried out when CP was on an Enrichment Placement at The Alan Turing Institute. CP thanks the members of the Turing and fellow Enrichment students for a stimulating experience. CP also thanks the participants of the 2025 Beg Rohu summer school of physics, in particular Elena Agliari, for discussion, and LPENS for support. CP is further supported by the UKRI AIMLAC CDT EP/S023992/1.
GA and BL are supported by STFC Consolidated Grant ST/T000813/1. 
BL is further supported by the UKRI EPSRC ExCALIBUR ExaTEPP project EP/X017168/1.
\\
We acknowledge the support of the Supercomputing Wales project, which is part-funded by the European Regional Development Fund (ERDF) via Welsh Government.

\noindent
{\bf Research Data and Code Access} --
The code and data used for this manuscript are available in Ref.~\cite{park_2025_17046571}. The repository is based on the template provided in Ref.~\cite{bennettTELOSCollaborationApproach2025}.

\noindent
{\bf Open Access Statement} -- For the purpose of open access, the authors have applied a Creative Commons Attribution (CC BY) licence to any Author Accepted Manuscript version arising.

\appendix
\renewcommand{\theequation}{\Alph{section}.\arabic{equation}}

\section{Duality in feature space}
\label{sec:duality}

In Sec.~\ref{sec:disordered}, we claimed that the output of a neural network is a linear combination of learned features. Here, we develop this viewpoint somewhat further. 

Using the notation of Sec.~\ref{sec:disordered}, 
\begin{align}
 \hat{y}_{i\alpha} = \sum_{j=1}^{n_{L-1}} W_{ij}^{(L)} \phi_{j\alpha},
 \qqquad
 \phi_{j\alpha} = \phi \left( z^{(L-1)}_{j} \left(x_{\alpha} \right)\right),
\end{align}
we note that the $\phi_{j\alpha}$'s are $n_{L-1} \times |\cD|$-dimensional matrices. These can be interpreted in two ways, namely as a set of $n_{L-1}$ vectors of size $|\cD|$ or as a set of $|\cD|$ vectors of size $n_{L-1}$, i.e.,
\begin{align}
 \phi_j \in \left\{ \phi_i \, \Big| \, \phi_i \in \mathbb{R}^{|D|}, i\in \left[1, n_{L-1}\right]\right\}
 \qquad \mbox{or} \qquad
  \phi_{\alpha} \in \left\{ \phi_{\beta} \Big| \phi_{\beta} \in \mathbb{R}^{n_{L-1}}, \beta \in \left[1, |\cD| \right] \right\}.
\end{align}
In the first interpretation, the output of a neural network is a linear combination of features, 
\begin{align}
 \hat{y}_{i} = \sum_{j=1}^{n_{L-1}} W_{ij}^{(L)} \phi_j, 
 \qqquad
 \hat{y}_i \in \mathbb{R}^{|\cD|},
\end{align}
whereas in the second interpretation, the output $\hat{y}_{\alpha} \in \mathbb{R}^{n_{L}}$ is a vector in the image given by the map $W^{(L)}$, which is not necessarily bijective, depending on the sizes of $n_L$ and $n_{L-1}$.

We may carry this distinction through to the loss function (\ref{eq:loss}), since the sums over the data index and node indices are independent. 
In the first case, we can write 
\begin{align}
 \cL = \frac{1}{|\cD|} \sum_{\alpha=1}^{|\cD|} \cH_{\alpha}, 
 \qqquad 
 \cH_{\alpha} \equiv \frac{1}{2}\sum_{i,j=1}^{n_{L-1}} J_{ij} \phi_{i\alpha} \phi_{j\alpha} - \sum_{i=1}^{n_{L-1}} h_{i\alpha} \phi_{i\alpha},
\end{align}
 where the term $\cH_{\alpha}$ is a random field spin-glass Hamiltonian of $n_{L-1}$ soft spins.
In the second case, we can define a local Hamiltonian along the node index direction, for fixed indices $i,j$,
\begin{align}
  \cL = \sum_{i,j=1}^{n_{L-1}} \cH_{ij}, \
  \qqquad
  \cH_{ij} \equiv \frac{1}{|\cD|} \sum_{\alpha=1}^{|\cD|} \left( \frac{1}{2} J_{ij} \phi_{i\alpha} \phi_{j\alpha} - h_{i\alpha} \phi_{j\alpha} \delta_{ij}\right)
\end{align}
which is a local Hamiltonian of a random field with $n = |\mathcal{D}|$ components and constant $J_{ij}$ for fixed $i$ and $j$.
Similarly, we can define the overlap or alignment $\bra h\phi\ket$ by contracting over either the data index or the node index, and these two interpretations capture different physics.
    
In the main part of the paper, we follow the first interpretation and consider the output as a linear combination of features $\phi_j$.

\section{Stochastic equation for the average level spacing}
\label{sec:derivation}

Let $x_i$ be an eigenvalue of a symmetric positive definite $N\times N$ matrix $X$ subject to stochastic dynamics,
\begin{align}
 \dot{x}_i = K_i + T \sum_{j \neq i} \frac{V_{ij}}{x_i - x_j} + \sqrt{T V_{ii}}\, \eta_i,
    \qqquad
    \eta_i \sim \cN(0, 1).
\end{align}
We assume the eigenvalues are ordered, $0\leq x_1<x_2<\ldots< x_N$. We are interested in the eigenvalue spacing, $S_i=x_{i+1}-x_i$, and its average value for each realisation,
\be
S = \frac{1}{N-1}\sum_{i=1}^{N-1} S_i = \frac{1}{N-1}\left(x_N-x_1\right).
\ee
In addition, one may average over an ensemble of trajectories. 

Our aim is to introduce a ``one-particle theory" for $S$, in the spirit of Ref.~\cite{PhysRevLett.95.246101}.
Subtracting the equation for the smallest eigenvalue $x_1$ from the one for $x_N$, one obtains
\begin{align}
    \dot{x}_N - \dot{x}_1
    = (N-1) \dot{S} 
    &=  K_N - K_1
    + T \left( \frac{V_{N1}}{x_N - x_1} - \frac{V_{1N}}{x_{1} - x_N} \right)
    \nn\\
    &
    + T \sum_{j=2}^{N-1} \left( \frac{V_{Nj}}{x_N - x_j} - \frac{V_{1j}}{x_1 - x_j} \right)
    + \sqrt{T V_{NN}} \, \eta_{N} - \sqrt{T V_{11}} \, \eta_{1}.
  \end{align} 
The noise term is the difference of two independent Gaussian random variables with zero mean and variances $T V_{N N}$ and $T V_{11}$, respectively. These can be combined as
  \begin{align}
    \sqrt{TV_{N N}} \, \eta_{N}
    - \sqrt{TV_{11}} \, \eta_{1}
    = \sqrt{T(V_{N N} + V_{11})} \, \eta,
    \qqquad
    \eta \sim \cN(0, 1).
  \end{align}
The Coulomb terms are non-trivial. We propose to replace the level spacings by the appropriately scaled average level spacing, i.e.,
\be
x_N-x_j = (N-j)S, \qqquad x_1-x_j = -(j-1)S.
\ee
We then write the combination of Coulomb terms as
\be
\frac{T}{N-1} \left( \frac{V_{N1}}{x_N - x_1} - \frac{V_{1N}}{x_{1} - x_N} \right)
+
\frac{T}{N-1} \sum_{j=2}^{N-1} \left( \frac{V_{Nj}}{x_N - x_j} - \frac{V_{1j}}{x_1 - x_j} \right)
=
V_S\frac{T}{S},
\ee
with
\be
\label{eq:VS}
V_S = \frac{2V_{N1}}{(N-1)^2} + 
\frac{1}{N-1} \sum_{k=1}^{N-2} \frac{1}{k} \left( V_{N,N-k} + V_{1,k+1}\right).
\ee
Here we used that $V_{ij}$ is symmetric and relabelled the summation index ($N-j=k, j-1=k$) in the two sums. 
The equation for $\dot S$ then takes the elegant form
\be
\dot{S} = K_S + V_S\frac{T}{S} + \sqrt{2TD_S}\, \eta,
\ee
where we also introduced the drift and the diffusion coefficient,
\begin{align}
\label{eq:KS}
K_S = \frac{1}{N-1}\left(K_N-K_1\right), \qqquad
D_S =  \frac{V_{11}+V_{NN}}{2(N-1)^2}.
\end{align}
In the main text, it is shown that the phase boundary between the high-temperature phase and the other phases depends on the ratio $K_S/V_S$. We determine this ratio as follows. We start with the gradient after training, $K_{ij}$ at $t=t_f$, for a given training run. This matrix is diagonalised, yielding the eigenvalues $K_i$. From this, we immediately obtain $K_S$ as the normalised difference between the largest and smallest eigenvalue, see Eq.~(\ref{eq:KS}). 
Using the same transformation, we also rotate $V_{ij}=\V[K_{ij}]$ to compute $V_S$ and $D_S$, see Eqs.~(\ref{eq:VS}, \ref{eq:KS}). Subsequently, we compute $K_S/V_S$ and compare this to LHS of Eq.~(\ref{eq:convergence_boundary}), to test the inequality.



\providecommand{\href}[2]{#2}\begingroup\raggedright\endgroup

\end{document}